\documentclass{article}
\usepackage{PRIMEarxiv}

\usepackage[utf8]{inputenc} 
\usepackage[T1]{fontenc}    
\usepackage{hyperref}       
\usepackage{url}            
\usepackage{booktabs}       
\usepackage{amsfonts}       
\usepackage{nicefrac}       
\usepackage{microtype}      
\usepackage{lipsum}
\usepackage{fancyhdr}       
\usepackage{graphicx}       
\usepackage{listings}
\usepackage{tcolorbox}
\usepackage{xcolor}
\usepackage{float}
\usepackage{pifont}
\usepackage{arydshln}
\newcommand{\xmark}{\ding{55}}%
\usepackage{tikz}
\usepackage{naive-ebnf}
\usepackage{caption}
\def\cmark{\tikz\draw[scale=0.30,fill=black](0,.35) -- (.15,0) -- (0.8,.7) -- (.15,.15);}
\def\cxmark{\tikz\draw[scale=0.30,fill=black](0,.35) -- (.15,0) -- (0.8,.7) -- (.15,.15) -- cycle (0.75,0.2) -- (0.67,0.2)  -- (0.48,0.7) -- cycle;}
\graphicspath{{media/}}     

\tcbset{
	userbox/.style={
		colback=blue!10!white,
		colframe=blue!50!black,
		sharp corners,
		boxrule=0.5mm,
		width=0.7\linewidth,
		left=0.3\linewidth
	},
	chatbotbox/.style={
		colback=gray!10!white,
		colframe=gray!50!black,
		sharp corners,
		boxrule=0.5mm,
		width=0.7\linewidth
	}
}

\title{SonicRAG : High Fidelity Sound Effects Synthesis based on Retrieval Augmented
	Generation 
}

\author{
  Yu-Ren Guo, Wen-Kai Tai\\
  Dept. of Computer Science and Information Engineering\\
  National Taiwan University of Science and Technology \\
  \texttt{{m11115115,wktai}@mail.ntust.edu.tw} \\
}

\begin{document}
\maketitle

\begin{abstract}
Large Language Models (LLMs) have demonstrated remarkable capabilities in natural language processing (NLP) and multimodal learning, with successful applications in text generation and speech synthesis, enabling a deeper understanding and generation of multimodal content. In the field of sound effects (SFX) generation, LLMs have been leveraged to orchestrate multiple models for audio synthesis. However, due to the scarcity of annotated datasets, and the complexity of temproal modeling. current SFX generation techniques still fall short in achieving high-fidelity audio. To address these limitations, this paper introduces a novel framework that integrates LLMs with existing sound effect databases, allowing for the retrieval, recombination, and synthesis of audio based on user requirements. By leveraging this approach, we enhance the diversity and quality of generated sound effects while eliminating the need for additional recording costs, offering a flexible and efficient solution for sound design and application.
\end{abstract}


\section{Introduction}
Large language models (LLMs) have become an integral part of modern life, demonstrating exceptional performance in text-based tasks across various fields, such as natural language processing (NLP) \cite{Raffel2019ExploringTL,Yang2023HarnessingTP} and expert systems. In the field of speech synthesis, models like VITS\cite{Kim2021ConditionalVA} have achieved remarkable success. However, sound effects (SFX) synthesis remains a challenging problem due to:
\begin{itemize}
	\item The lack of efficient and precise annotation methods for sound events.
	\item The complexity of modeling multiple overlapping sound events occurring simultaneously.
\end{itemize}

While LLMs can analyze textual descriptions to extract sound event sequences and break down tasks for sound synthesis \cite{Huang2023AudioGPTUA,Wang2024AudioAgentLL}, the quality of audio generated by latent diffusion models (LDMs) such as AudioLDM still falls short of professionally recorded studio audio. Additionally, due to the nature of audio, humans are challenging to edit and reconstruct details as they wish in image editing. 

To address these challenges, this paper proposes \textbf{SonicRAG}, integrates retrieval-augmented generation (RAG) \cite{Lewis2020RetrievalAugmentedGF} with LLMs' ability to analyze complex descriptions. By leveraging a small set of high-quality sound effects and describing sound tracks with abstract language—\textbf{Mixer Script}, our approach enables language models to synthesize diverse and user-intended sound effects, bridging the gap between AI-generated audio and professional-quality recordings.

\section{Releated Work}
\subsection{Audio Generation}
Depending on the characteristics of different sounds, the suitable sound synthesis methods may vary. For example, in the field of speech synthesis, VAE-based models such as VITS\cite{Kim2021ConditionalVA} have achieved excellent performance in text-to-speech tasks. In music generation, Transformer-based models like AudioCraft\cite{copet2023simple} and MuseNet\cite{openai2019musenet} have also produced promising results. Meanwhile, in sound effect generation, the current mainstream approach involves diffusion-based models, such as AudioLDM\cite{Liu2023AudioLDMTG} and Stable Audio\cite{Evans2024StableAO}. Although these methods are capable of generating content that aligns well with textual descriptions, they tend to struggle when dealing with more abstract or nuanced creative requirements.

\subsection{Assistants Based on LLMs}
In the era of AI-generated content (AIGC), many enterprises have demonstrated that large language models (LLMs) based on the Transformer architecture are capable of understanding and assisting in meeting human needs. Such as ChatGPT, DeepSeek \cite{Achiam2023GPT4TR,DeepSeekAI2025DeepSeekR1IR}help solve problems through conversational question-and-answer interactions. GitHub Copilot, powered by OpenAI Codex \cite{Chen2021EvaluatingLL}, analyzes engineers' code and requirements to derive solutions. Additionally, Microsoft Copilot \cite{MicrosoftCopilot}integrates LLMs with Microsoft Graph, further enhancing the user experience of office software. These applications prove that LLMs can serve as central decision-making hubs capable of analyzing multimodal inputs and generating instructions to accomplish a wide variety of tasks.

\subsection{Prompt Engineering}
Since LLMs typically contain billions of parameters, fine-tuning them for every specialized task would require enormous computational resources. Fortunately, the self-attention mechanism of LLMs allows them to infer user intent based on contextual information. This enables the use of prompt engineering techniques to achieve strong performance without the need for retraining. Brown \emph{et al.} \cite{Brown2020LanguageMA} proposed that when a model has a sufficiently large number of parameters, providing it with a small number of task examples can guide it toward the desired outcome. Additionally, Wei \emph{et al.} \cite{Wei2022ChainOT} suggested that prompts can be designed to simulate human thinking processes, further enhance the model’s performance on specialized tasks.

\subsection{Retrieval-Augmented Generation }
Although LLMs demonstrate the ability to retain knowledge, it remains infeasible to train an all-knowing model within a limited budget. Lewis \emph{et al.} \cite{Lewis2020RetrievalAugmentedGF}  shows that parametric and non-parametric memory doing postive interaction on knowledge-intensive tasks. This means that providing pre-query information along with the target prompts can improve the performance of LLMs. In our framework, LLMs analyze user's requirement, retrieval the audio assets, and generate audio sequence for synthesis sound effects.

\section{Methodology}
\subsection{Mixer Script}
\begin{figure}[ht]
	\centering
	\includegraphics[width=0.7\textwidth]{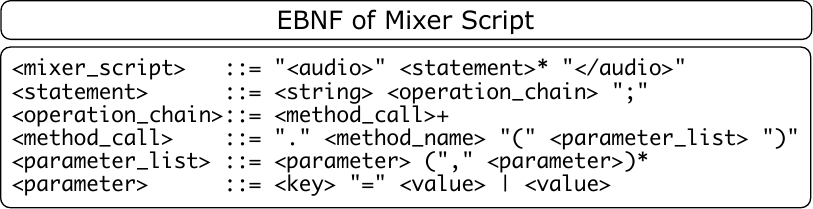}
	\caption{\emph{Extended Backus-Naur Form of Mixer Script }}
	\label{fig:ebnf}
\end{figure}

\begin{figure}[H]
	\centering
	\includegraphics[width=0.8\textwidth]{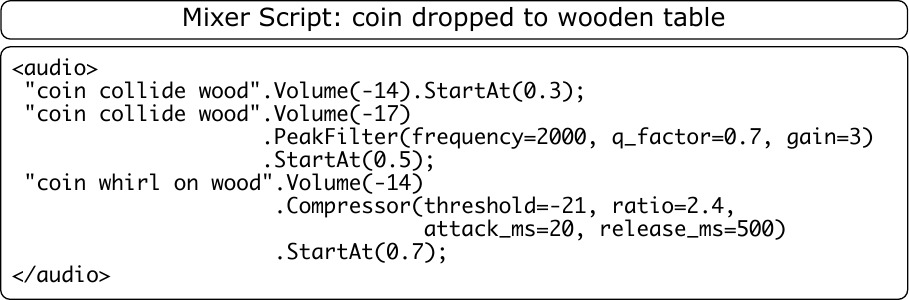}
	\caption{\emph{The sound of "coin dropped to wooden table" descriped by Mixer Script.}}
	\label{fig:mxs}
\end{figure}

\begin{figure}[H]
	\centering
	\includegraphics[width=0.8\textwidth]{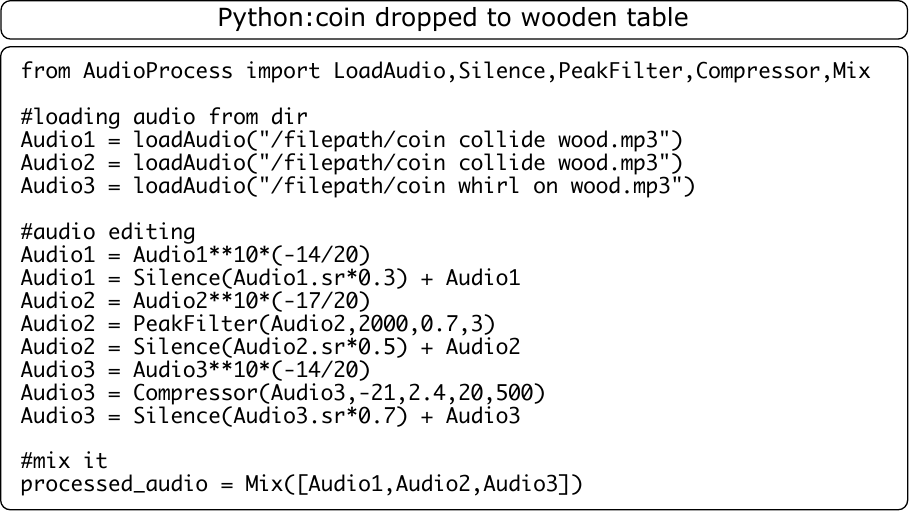}
	\captionsetup{width=0.75\textwidth}
	\caption{\emph{The python code of "coin dropped to wooden table", using pre define library "AudioProcess" for signal processing.}
	\label{fig:py}}
\end{figure}

Although LLMs can generate code for synthesizing sounds and execute the code to obtain audio, for sound designers and mixing engineers,  understanding an unfamiliar programming language can be a significant hurdle. Moreover, when generating complex scripts, LLMs are prone to producing code that fails to execute reliably due to model limitations. To address this, we design an abstract scripting language—Mixer Script, which encapsulates audio signal processing methods. By leveraging method chaining to describe the parameters of each audio asset in a task, our approach reduces the difficulty of reading generated code and simplify the complexity of code generation for LLMs. The syntax of  Mixer Script is defined in Figure \ref{fig:ebnf} using EBNF (Extended Backus-Naur Form) \cite{ISO14977}, and the supported audio processing methods are defined in Table \ref{tab:provided}, As an example, the script for sound synthesis "coin dropped onto wooden table" generated by using \textbf{Mixer Script} and\textbf{ Python}, is used to compare readability between Figure \ref{fig:mxs} and Figure \ref{fig:py}.

\begin{table}[htbp]
	\centering
	\begin{tabular}{lll}
		\toprule
		Mehtods & Parameters & Adjusts \\
		\midrule
		Volume & targetLUFS & volume to target loudness \\
		Compressor & threshold, ratio, attack\_ms, release\_ms & dynamic of sound\\
		Reverb & room\_size, dry\_wet & adding the Reverb Effect \\
		PeakFilter & frequency, q\_facter, gain & voulme of specified frequency \\
		LowPassFilter & frequency & cutoff volume over target frequency\\
		HighPassFilter & frequency & cutoff volume below target frequency\\
		StartAt & at & start time of sound event \\
		StopAt & at, fade\_out\_duration & fade out the sound event \\
		\bottomrule
	\end{tabular}
	\caption{Audio Processing Methods supported by MixerScript}
	\label{tab:provided}
\end{table}

%

%
\subsection{Metadata of Sound Events}
\begin{figure}[H]
	\centering
	\includegraphics[width=0.30\textwidth]{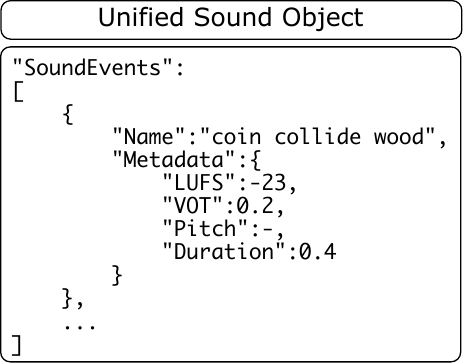}
	\captionsetup{width=0.75\textwidth}
	\caption{\emph{Example of unified sound object. In the sound event list, each sound would be packaged to a object with it's name  and metadata.}}
	\label{fig:mose}
\end{figure}

While LLMs pretrained with multimodal have the ability to process audio tokens, but most that are optimized for speech and exhibit limitations in understanding the characteristics of sound events. For example, they may struggle to interpret pitch or complex acoustic scenes. Therefore, it becomes necessary to use textual representations to help LLMs comprehend the metadata of sound events.
In our framework, we identify four key parameters essential to sound design, namely \textbf{Loudness}, \textbf{Voice onset time}, \textbf{Pitch}, and \textbf{Duration}. By packaging this information together with a textual description into a unified sound object, we enable large language models to infer the characteristics of corresponding sound. This design also allows our framework to leverage LLMs trained solely on textual data for sound generation tasks. Figure \ref{fig:mose} is an example of sound object.

\subsection{Context-Aware Retrieval}
In Retrieval-Augmented Generation (RAG), prompts serve not only to drive LLMs but also to retrieve essential reference data. However, in the task of sound sequence synthesis, each user input does not necessarily require  retrieving for new sound assets, such as re-synthesis or fine-tuning may not demand additional queries. While we could ignore this and perform retrieval regardless, excessive and irrelevant context could negatively impact the performance of LLMs.  

To overcome this, we have implemented a mechanism that allows LLMs to autonomously analyze the dialogue context and determine whether new sound assets are needed. This approach offers two key advantages:  
(1) Sparing complex rule-based scripts to analyze each user input.  
(2) It allows query terms to be formulated in a specified language, preventing cross-language queries from reducing the accuracy of the vector database.

\subsection{Framework}
In our proposed SonicRAG framework, the generation of an audio synthesis script $\mathcal{S}$ is modeled as a Markov decision process \cite{bellman1957dp}:

\begin{equation}
	\mathrm{SonicRAG} = P(\mathcal{S}_t \mid \mathcal{S}_{t-1},\mathcal{P},\mathcal{M}, \mathcal{L},\mathcal{R}),
\end{equation}

Where $\mathcal{P}$ is the user prompt describing the desired sound, $\mathcal{M}$ represents predefined audio processing methods, $\mathcal{L}$ is language model which is used to drive the system, and $\mathcal{R}$ consists of retrieved audio assets, which is modeled as:
 
\begin{equation}
	\mathrm{Retrieval} = P(\mathcal{R}_t \mid \mathcal{R}_{t-1},\mathcal{P}, \mathcal{L}),
\end{equation}
 
It means that whether retrieval is updated depends on current state and the LLM's analysis of prompt. The script generation process is context-aware, incorporating both retrieval-augmented information and domain-specific synthesis methods to iterate listening experience and relevance of the generated sounds.


\section{Experiments}
\subsection{Setup}
For our experiments, we selected gpt-4o-audio-preview as the engine to drive SonicRAG. This large language model supports audio as input modality, meaning it can process audio tokens directly. Without relying on speech recognition, users can describe the desired sound effects using spoken descriptions, including onomatopoeia. We recorded a series of coin-related sound with sampling format of 48kHz/24bit for assets and named the files according to their corresponding events, the filenames are also stored in a vector database to facilitate retrieval based on specific queries.

\subsection{Case study: Sound Design Workflow}
Sound design plays a crucial role in multimedia industries such as film and game design. Well-crafted sound effects enhance immersion, elevating the overall experience. For sound designers who have already built extensive sound libraries, the key challenge is efficiently analyzing creative intent and transforming imagination into high-quality sound effects.

SonicRAG addresses this need by enabling interactive dialogue, allowing for iterative refinement of the designer’s vision and seamless integration of existing assets into a cohesive output. The following demonstrates SonicRAG in real-world applications:
\begin{figure}[!ht]
	\centering
	\includegraphics[width=1\textwidth]{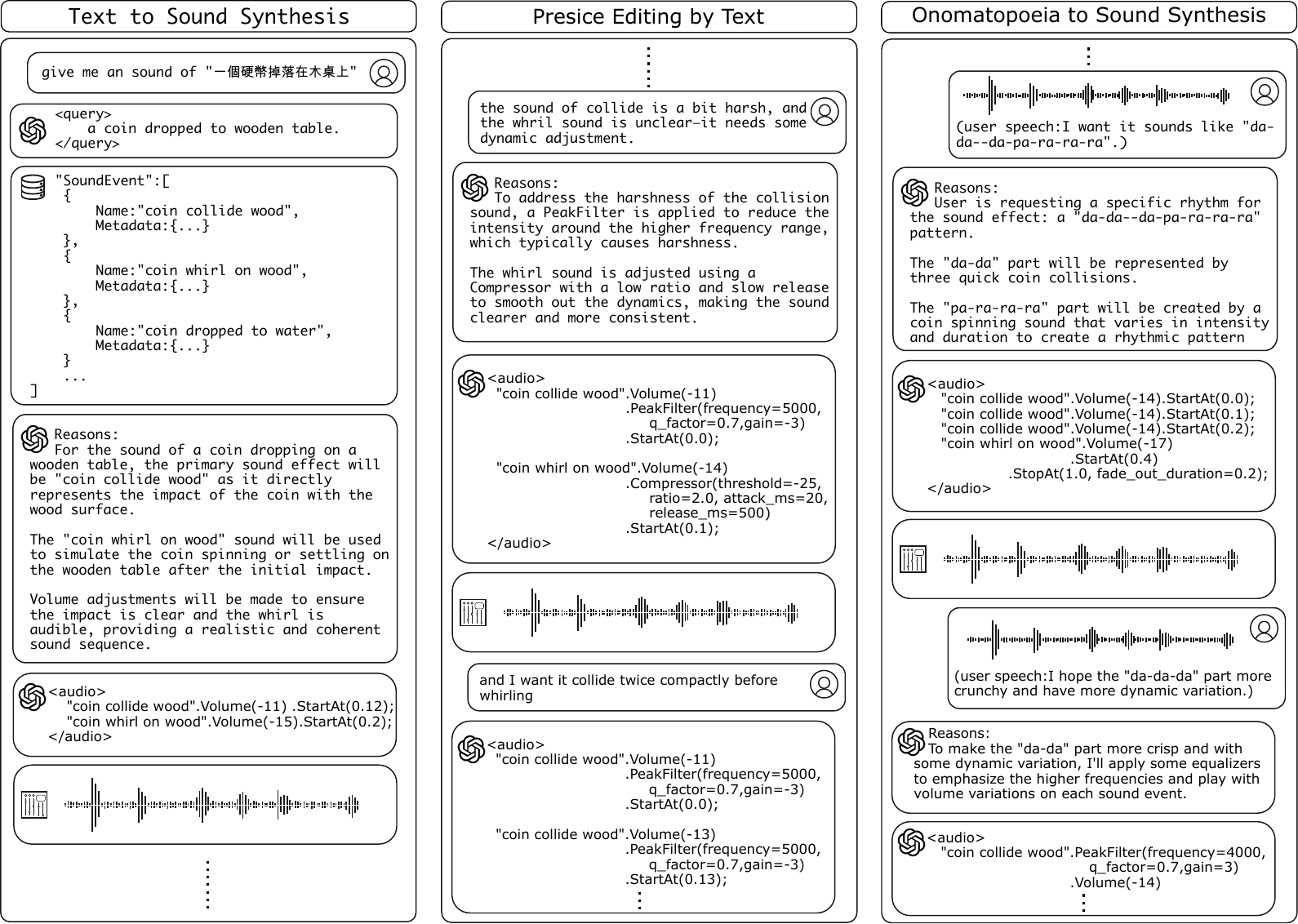}
	\captionsetup{width=0.85\textwidth}
	\caption{\emph{Behavior of SonicRAG in diffrent working situation. Left side shows how is RAG and Mixer enable LLM to synthsis audio. Central depicts adjustability of framework. Right part exhibits multimodel LLM can understand onomatopoeias and represent it to other media.}}
	\label{fig:behavior}
\end{figure} 

\subsection{Comparision}
SonicRAG can reuse pre-recorded sound assets and synthesize sound effects through Mixer Script, giving it an advantage in both sound quality and precise sound adjustment. Additionally, by leveraging multimodal training, LLMs can be further extended to achieve speech-to-sound and even onomatopoeia-to-sound generation. 

Due to the limitations of neural network models, previous approaches had to downsample the signal to around 16kHz or use spectrogram-based training and generation.While this yields satisfactory results for speech synthesis, the loss of details caused by aliasing becomes more apparent in sound effects. Precise editing means user could adjust components detail (e.g. Frequency, Dynamics and Voice onset time) of sound. Audio generation based on Latent Diffusion Models remains difficult to control. Although LLMs can interpret onomatopoeias and abstract concepts, they still struggle to consistently produce the desired results when applied to systems like WavCraft and WavJourney. Table \ref{tab:comparison} compares SonicRAG with existing approaches.

\begin{table}[!ht]
	\centering
	\renewcommand{\arraystretch}{1.3} 
	\setlength{\tabcolsep}{10pt} 
	\begin{tabular}{lccc} 
		\toprule
		Method & Native sampling & Precise editing by text & onomatopoeia to sound\\
		\midrule
		AudioGen \cite{copet2023simple} &\xmark & \xmark & \xmark  \\
		AudioLDM \cite{Liu2023AudioLDMTG} & \xmark & \xmark & \xmark \\
		WavCraft \cite{Liang2024WavCraftAE} &\xmark & \cxmark & \cxmark \\
		WavJourney  \cite{Liu2023WavJourneyCA}  &\xmark & \cxmark & \cxmark \\
		SonicRAG$_{(ours)}$  &\cmark & \cmark & \cmark \\
		\bottomrule
	\end{tabular}
	\caption{\emph{Comparison of capabilities between SonicRAG and recent audio generation methods.}}
	\label{tab:comparison}
\end{table}

\subsection{Specialized Generating}
In this section, we present a collection of synthesis prompts related to coin sounds, such as “Coin dropped onto a wooden table” and “Multiple coins jingling together in a person’s hand”. We compare the results generated by our method with those from previous approaches using several evaluation metrics:
Fr{\'e}chet Audio Distance (FAD) \cite{Kilgour2018FrchetAD}, It measures the similarity between generated and real audio distributions based on deep feature embeddings, and we use ARCA23K\cite{Iqbal2021} for the ground truth set. 
Contrastive Language-Audio Pretraining Score (CLAP)\cite{elizalde2022claplearningaudioconcepts,xiao2024CLAPScore}, evulates semantic correlation between natural laguage and audio. 
Signal-to-Noise Ratio (SNR), a traditional audio quality metric that quantifies the amount of desired signal relative to background noise. We report the average comparison results for each method based on this metric.

\begin{table}[htbp]  
	\centering  
	\setlength{\tabcolsep}{10pt}
	\renewcommand{\arraystretch}{1.3} 
	\begin{tabular}{lrrr}    
		\toprule    Methods & FAD $\downarrow$ & CLAP $\uparrow$ & SNR $\uparrow$\\    
		\midrule
		AudioGen \cite{copet2023simple} & 32.1  & 0.48  & 29.1 \\        
		AudioLDM \cite{Liu2023AudioLDMTG} & 22.3  & 0.44  & 40.4 \\    
		StableAudio \cite{Evans2024StableAO} & \textbf{21.4}  & 0.56  & 37.0\\    
		WavCraft \cite{Liang2024WavCraftAE}& 25.8  & 0.32  & 53.8 \\    
		WavJourney \cite{Liu2023WavJourneyCA} & 27.9  & 0.44  & 40.8 \\    
		\hdashline
		SonicRAG$_{(ours)}$ & 25.5  & \textbf{ 0.67}  & \textbf{ 88.2} \\ 
		\bottomrule   
	\end{tabular}
	\caption{\emph{Evaluation results on Specialized Generating}}   
	\label{tab:score}
\end{table}


\section{Conclusion}
\subsection{Summary}
In this paper, we proposed SonicRAG, a retrieval-augmented generation (RAG) framework tailored for sound synthesis. Unlike conventional methods that rely solely on pre-trained generative models or manual sound retrieval, SonicRAG enables an interactive workflow where users provide natural language prompts, and the system dynamically retrieves and mixes audio assets to generate high-quality sound effects. 

Our experimental results demonstrate that SonicRAG achieves superior performance in requirement accuracy and synthesis flexibility compared to existing methods. By leveraging Mixer Script for structured audio translation, we provide sound designers with greater control over the final output while maintaining ease of use. Furthermore, our approach reduces the reliance on extensive retraining, making it a scalable solution for real-world applications.

Overall, SonicRAG presents a novel and practical approach to AI-assisted sound design, bridging the gap between retrieval-based and generative methodologies while offering an intuitive interaction model for creative professionals.

\subsection{Limitation}
SonicRAG provide user an creative and interactive interface for synthesis high fidelity sound effect. But it still have limitations: 1): Infrastructure: the framework needs user already have text-relational audio files for assets. 2): Perception: Due to different sense of hearing, capability of effects provided by Mixer Script may not enough for everyone. 3): Efficiency: it require larger size LLMs such as GPT-4o for comprehend abstract tasks. 

\subsection{Future Work}
Use different sounds to ensemble euphonious results is a complex problem that encompasses multiple disciplines like SEC (sound event classfication), SED (sound event detection with timestamp) and ASC (Acoustic sence classification). It isn't solved effectively in textual-based methods so far. Future research could explore using task-specific pre-trained embeddings to assist in fine-tuning LLMs, optimizing their performance for sound synthesis.

\bibliographystyle{unsrt}

\end{document}